\documentclass[english,times,pra,twocolumn,superscriptaddress]{revtex4-2}
\usepackage{bm}
\usepackage{amsmath}
\usepackage{amssymb}
\usepackage{graphicx}
\usepackage{newtx}

\begin{document}
\title{Displaced photon helicity states: a quantum realization of Arago-Fresnel
interference}
\author{Yan Zhang}
\affiliation{Institute of Applied Physics and Materials Engineering, University
of Macau, Macao, S.A.R., China}
\author{Xiang Liu}
\affiliation{Institute of Applied Physics and Materials Engineering, University
of Macau, Macao, S.A.R., China}
\author{Hou Ian}
\email{houian@um.edu.mo}

\affiliation{Institute of Applied Physics and Materials Engineering, University
of Macau, Macao, S.A.R., China}
\begin{abstract}
The laws of Arago and Fresnel relate the polarization of light to
its capacity for interference. We investigate their quantum counterpart
at the level of the fundamental helicity states of a single photon.
To this end, we introduce displaced photon helicity states (DPHS),
constructed by coherently displacing single-photon helicity eigenstates.
The displacement bridges the microscopic spin degree of freedom and
macroscopic polarization observables, enabling interference effects
that would otherwise be inaccessible to isolated photons. We derive
analytical expressions for the Stokes parameters associated with the
interference between DPHS and coherent optical fields and show that
opposite helicities generate distinct parameter dependences in the
resulting fringe structure. The interference constitutes a quantum
realization of Arago-Fresnel interference and permits the helicity
of an unknown photon to be inferred from the Stokes response of the
combined field. Beyond this application, DPHS generalize the framework
of displaced nonclassical states from photon-number excitations to
the helicity degree of freedom of light.
\end{abstract}
\maketitle

\section{Introduction}

The relation between interference and polarization was historically
first established through the four laws of Arago and Fresnel~\citep{Arago1811,AragoFresnel},
which specify the conditions under which two polarized light beams
produce observable interference~\citep{Crew_book,Buchwald_book}.
In modern terminology, the first two laws state that interference
is permitted for copolarized beams but suppressed for orthogonally
polarized ones, while the remaining laws establish the role of coherence
in restoring interference between beams sharing a common origin. Subsequent
formulation through the Stokes parameters~\citep{Stokes} provided
a quantitative description of these observations and completed the
classical theory of Arago-Fresnel (AF) interference.

Although polarization and coherence have now became foundational concepts
of modern optics~\citep{HanburyBrown,Glauber}, the AF laws remained
formulated at the level of classical interference of waves. Their
reinterpretation within microscopic theories emerged only much later.
In particular, stochastic approaches based on field correlations~\citep{Barakat}
have shown that the polarization dependence of AF interference can
be derived from random optical fields~\citep{Baraket_1983}, thereby
connecting the interference laws with coherence theory. Nevertheless,
the role of genuinely nonclassical states in AF interference has received
comparatively little attention.

Recent advances in quantum optics have led to extensive studies of
engineered nonclassical states~\citep{Dodonov_book}, including squeezed
states, Schrödinger cat states~\citep{Agarwal,Asavanant,Takase},
photon-added coherent states~\citep{Zavatta,Fadrny,Kala}, and displaced
Fock states~\citep{Boiteux,Kral,Othman}. Among these, displaced
Fock states combine the coherence introduced by displacement with
the nonclassical properties of Fock-state excitations. They have been
widely studied under various contexts ranging from state engineering~\citep{Lvovsky2002,Keil}
to metrology~\citep{Wolf} and entanglement generation~\citep{Biagi,LiMinghui}.
However, existing studies have focused primarily on photon-number
properties, whereas the influence of photon helicity on displaced-state
interference remains largely unexplored.

In this work, we consider \emph{displaced photon helicity states}
(DPHS) $D(\bm{\alpha})\left|\mathbf{1}_{\theta,\phi}\right\rangle $
constructed by applying the displacement operator to a single-photon
state expressed in the gauge-invariant helicity basis $\{\left|\circlearrowleft\right\rangle ,\left|\circlearrowright\right\rangle \}$~\citep{Fierz,Jauch_book}:
\begin{equation}
\left|\mathbf{1}_{\theta,\phi}\right\rangle =\cos\theta\left|\circlearrowleft\right\rangle +e^{i\phi}\sin\theta\left|\circlearrowright\right\rangle .\label{eq:arb_spin}
\end{equation}
Consequently, DPHS incorporate both the coherent amplitude associated
with displacement and the helicity degree of freedom associated with
the underlying single-photon state. This combination effectively amplifies
photonic helicities to detectable macroscopic interference. In particular,
we study the interference generated between a coherent state, i.e.
the displaced vacuum $D(\alpha)|0\rangle$, and a DPHS. Following
the traditional AF configuration, two beams carrying these two states
are projected onto a common polarization basis and subsequently recombined.
Inferring the fringe patterns through the Stokes parameters as in
the classical formulation, we nevetheless find that DPHS produces
additional $(\theta,\phi)$-dependent structures that are absent from
classical polarization interference.

Most notably, the Stokes parameter $S_{0}$ exhibit a quantum signature:
it acquires a helicity-dependent asymmetry with respect to $\theta=\pi/4$
that has no counterpart in classical AF laws. The present construction
therefore establishes a quantum realization of the AF polarization-interference
relation in terms of photonic helicity. In other words, helicity information
can be encoded into macroscopic interference through displacements.
The quantum AF interferometry thus provides not only a means to detect
photons without a single-photon detector, but also a purely optical
alternative for detecting photonic helicity over mechanical means
such as Ref.~\citep{Beth}.

The remainder of this paper is organized as follows. Section II introduces
the DPHS formalism. Section III furnishes the interferometric configuration
and derives the Stokes parameters for both the case of interfering
two coherent states as a baseline reference and of interfering a DPHS
with a coherent state. Section IV analyzes interference patterns,
compares them with the baseline, and discusses their implied relations
to the helicity of an incident photon. Conclusions are presented in
Sec. V.

\section{Displacing helicity state}

To construct a displaced helicity state of photons from the premise
Eq.~(\ref{eq:arb_spin}), we first consider a general Fock state.
Since Fock state is a superposition of tensor-product states accounting
for the symmetry about particle indistinguishability~\citep{Geroch_book},
we have the expression
\begin{align}
\left|\mathbf{n}_{\theta,\phi}\right\rangle  & =\left|\mathbf{1}_{\theta,\phi}\right\rangle ^{\otimes n}\nonumber \\
 & =\sum^{n}_{k=0}\binom{n}{k}e^{ik\phi}\cos^{n-k}\theta\sin^{k}\theta\left|\circlearrowleft^{n-k},\circlearrowright^{k}\right\rangle .\label{eq:Fock_state}
\end{align}
The second line follows from the binomial expansion of $n$-photon
tensor product such that each eigenvector $\left|\circlearrowleft^{n-k},\circlearrowright^{k}\right\rangle $
indicates a possible partition of $n$ photons into two categories,
associated respectively with the left- and the right-helicity. The
Bloch angles $\theta$ and $\phi$ of the orientation unit vector
$\mathbf{n}$ is entirely encoded in the coefficients.

From Eq.~(\ref{eq:Fock_state}), the coherent state of complex displacement
$\alpha$ can be generalized from the familiar definition to a vector
form depending on the Bloch angles as (Cf. App.~\ref{sec:appendix a}
for all the derivations in this section) 
\begin{multline}
\left|\bm{\alpha}_{\theta,\phi}\right\rangle =e^{-|\alpha|^{2}/2}\sum^{\infty}_{n=0}\sum^{n}_{k=0}\frac{\alpha^{n}e^{ik\phi}}{\sqrt{(n-k)!k!}}\\
\cos^{n-k}\theta\sin^{k}\theta\left|\circlearrowleft^{n-k},\circlearrowright^{k}\right\rangle ,\label{eq:cohere_state}
\end{multline}
which separates the displacements into the left- and the right-helicity
parts. Note that the addition of subscripts $\theta$ and $\phi$
emphasizes the displacement from vacuum be constructed from a specific
helicity mode as stipulated in Eq.~(\ref{eq:arb_spin}). For usual
contexts, these Bloch angles are irrelevant and become implicit: e.g.
a single-mode laser source is typically linearly polarized, implying
$\theta=\pi/4$ and $\phi=0$ and reducing Eq.~(\ref{eq:cohere_state})
to its familiar form without $\theta$ and $\phi$. Note that the
combinations of helicities implied in the Bloch angles are distinct
from the quadrature phase $\arg\alpha$ of the displacement. Therefore,
the detectable polarizations of an arbitrary coherent state, or a
DPHS therefrom, will depend on all three variables $\theta,\phi$
and $\arg\alpha$.

\begin{figure}
\includegraphics[bb=0bp 0bp 509bp 400bp,clip,width=9cm]{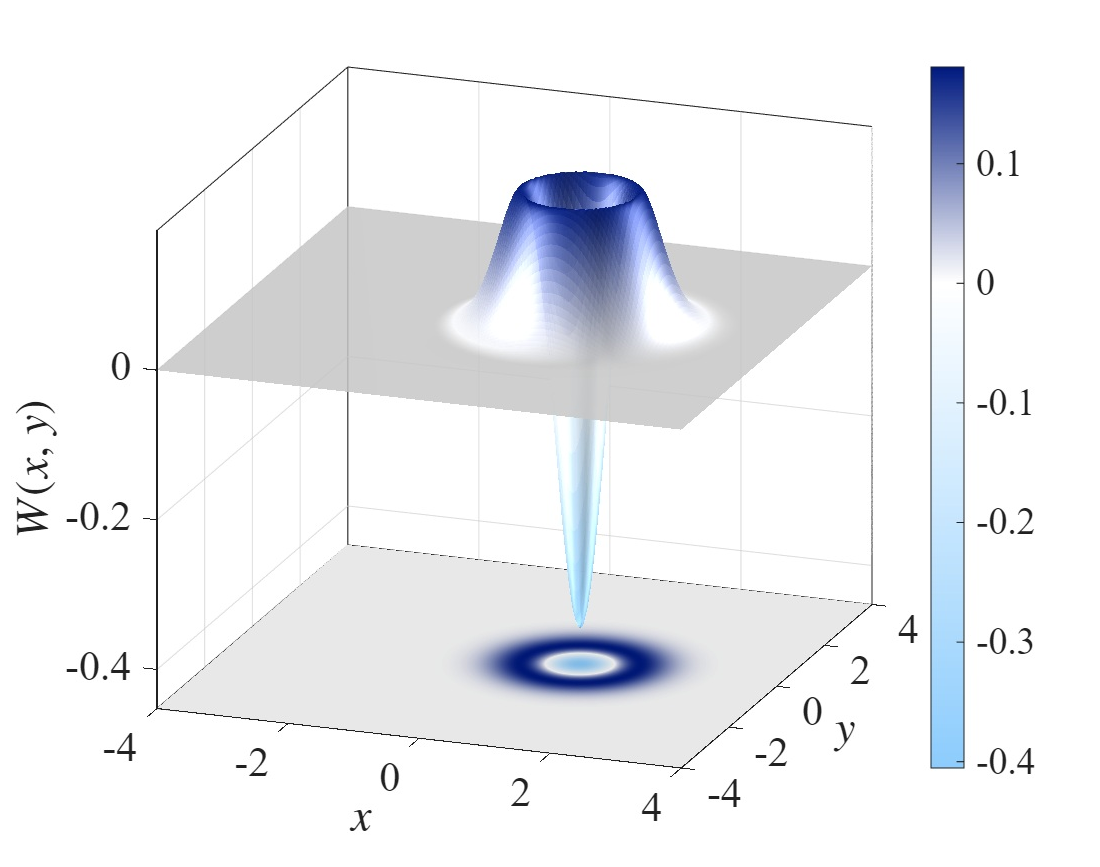}

\caption{Surface plot of the Wigner distribution $W(x,y)$ against the quadratures
$x$ and $y$ for a displaced helicity state of displacement $\alpha=1.0$
and Bloch angles $\theta=\pi/2$, $\phi=0$. The contour projected
on $xy$-plane is also plotted. The distribution remains identical
for displacing either from the left $\left|\circlearrowleft\right\rangle $
or the right $\left|\circlearrowright\right\rangle $ photon helicity.~}\label{fig:DSS_Wigner_plot}
\end{figure}

Consequently, analytic expressions for the left and the right DPHS,
serving as the basis vectors of superpositions for arbitrary DPHS
to be detected, can be computed from Eq.~(\ref{eq:cohere_state})
through the identities
\begin{align}
D(\bm{\alpha}_{\theta,\phi})\left|\circlearrowleft\right\rangle  & =(a^{\dag}-\alpha^{\ast}\cos\theta)\left|\bm{\alpha}_{\theta,\phi}\right\rangle ,\label{eq:left_DPHS}\\
D(\bm{\alpha}_{\theta,\phi})\left|\circlearrowright\right\rangle  & =(\bar{a}^{\dag}-\alpha^{\ast}e^{-i\phi}\sin\theta)\left|\bm{\alpha}_{\theta,\phi}\right\rangle ,\label{eq:right_DPHS}
\end{align}
where $\{a,a^{\dag}\}$ (resp. $\{\bar{a},\bar{a}^{\dag}\}$) denotes
the annihilation and creation operator pair for the left (resp. right)
photon helicity mode. Correspondingly, $D(\bm{\alpha}_{\theta,\phi})=\exp\{\alpha(a^{\dag}\cos\theta+\bar{a}^{\dag}e^{i\phi}\sin\theta)-\mathrm{h.c.}\}$
is the vector displacement operator of the displacement $\bm{\alpha}_{\theta,\phi}$
carrying the data of both the usual complex amplitude $\alpha=|\bm{\alpha}_{\theta,\phi}|$,
as well as the elliptical polarization angles $\theta$ and $\phi$.
The latter can be regarded respectively as the polar and the azimuthal
angles on a Bloch sphere. Experiment wise, the RHS of the identities
can be easily realized by applying a quadrature detection $q=a+a^{\dag}$
on the coherent state and removing the offset but either post-selecting
idler photons or mixing with a beam carrying displacement $(-\alpha)$~\citep{LiMinghui}.
Though distinct at appearance, the two displaced states cannot be
discriminated through their own measurements alone. For example, given
identical displacement $\alpha$, their Wigner distributions determined
by photo detector levels along varying quadrature angles will remain
the same. An exemplary Wigner plot is shown in Fig.~\ref{fig:DSS_Wigner_plot}.
Nevertheless, interfering these DPHS against a reference coherent
state in an AF interferometer will produce fringes discriminating
Eq.~(\ref{eq:left_DPHS}) from Eq.~(\ref{eq:right_DPHS}).

\section{Interferometry of photon helicity}

\subsection{Quantum Arago-Fresnel interferometer}

\begin{figure}
\includegraphics[bb=10bp 90bp 1150bp 750bp,clip,width=9cm]{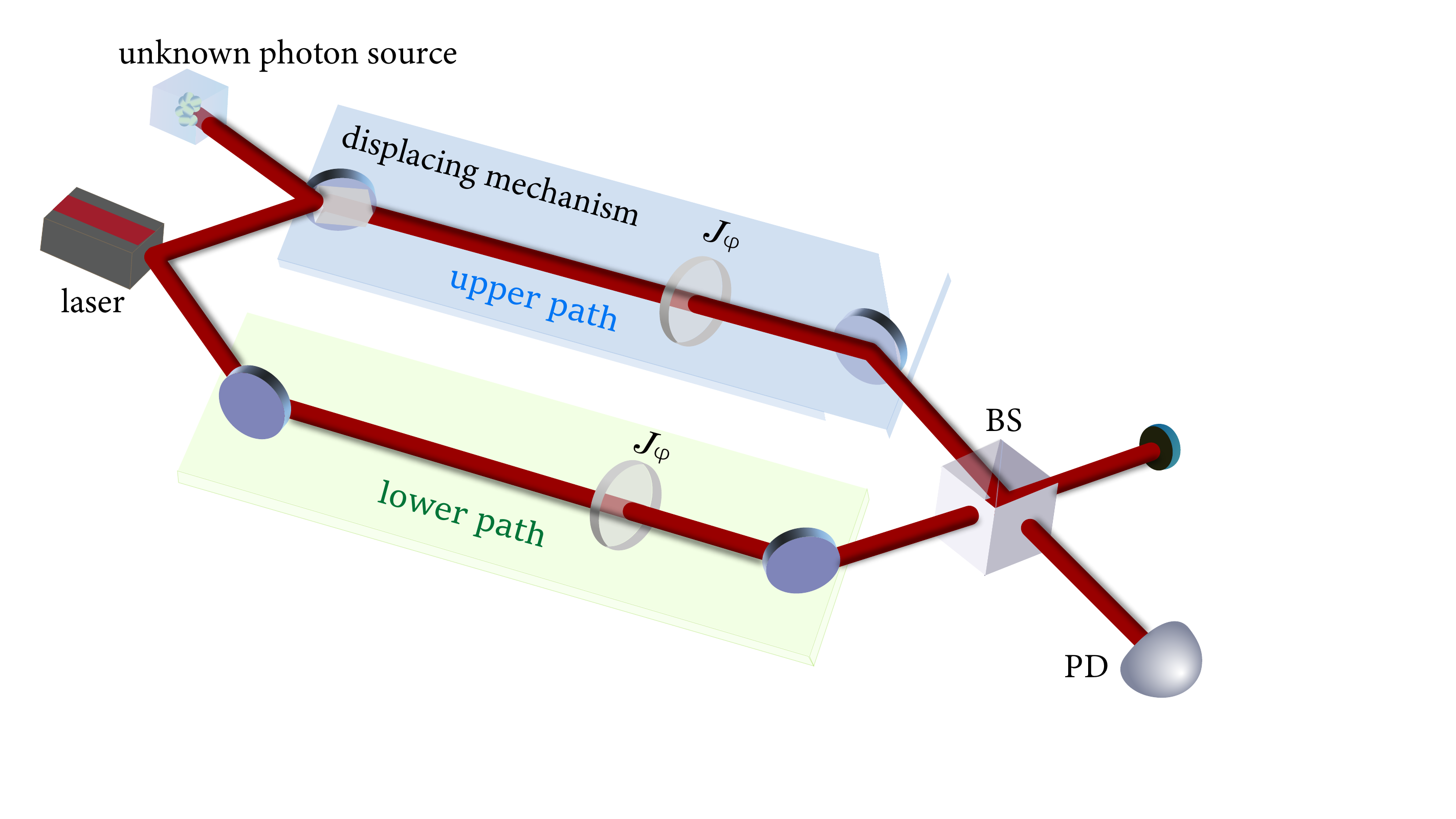}

\caption{Model setup of a quantum Arago-Fresnel interferometer. An incident
photon with unknown helicity is coherently displaced by a interfering
reference laser to form a detectable displaced helicity state. The
displacement is realized by parametric down conversion pumped by the
reference laser with a vacuum idler input~\citep{LiMinghui}: DPHS
comes out at the signal output when the idler output is projected
onto a single-photon detector.~}\label{fig:interferometer_model}
\end{figure}

The setup of a quantum Arago-Fresnel interferometer is shown in Fig.~\ref{fig:interferometer_model}.
Similar to classical setups, two incident beams are brought to the
same polarization plane to interfere via a beam splitter, whose balanced
output proportional to the zeroth Stokes parameter $S_{0}$ is sent
to a photo detector for quadrature measurement. The upper beam, produced
by parametric down converting a coherent pump beam with a photon of
unknown spin state, implements the algebraic effects of\emph{ }Eqs.~(\ref{eq:left_DPHS})-(\ref{eq:right_DPHS})
and generates a DPHS serving as the target signal. The lower beam,
also in a coherent state, serves as a reference. When the coherent
pump and the lower beam are derived from the same single-mode polarized
laser, the setup is reminiscent of a homodyne detection where the
reference beam plays the role of local oscillator.

The classical step of ``bringing into the same polarization
plane'' is realized by letting the two beams through identically oriented
linear polarizers, whose effect on a single photon is described by
the elementary Jones matrix
\begin{equation}
\hat{\mathcal{J}}=\frac{1}{2}\left[\left|\circlearrowleft\right\rangle \left\langle \circlearrowleft\right|+\left|\circlearrowright\right\rangle \left\langle \circlearrowright\right|+e^{2i\varphi}\left|\circlearrowright\right\rangle \left\langle \circlearrowleft\right|+e^{-2i\varphi}\left|\circlearrowleft\right\rangle \left\langle \circlearrowright\right|\right]\label{eq:Jones}
\end{equation}
in the photon-helicity basis with unity trace. The angle $\varphi$
denotes the polarizer orientation from the horizontal axis determined
by the polarizing direction of the reference beam at $\theta=\pi/4$
and $\phi=0$. Composing Eq.~(\ref{eq:Jones}) to obtain a multi-photo
Jones matrix $\hat{\mathcal{J}}^{\otimes n}$on the Fock state represented
by Eq.~(\ref{eq:Fock_state}) (Cf. App.~\ref{sec:appendix b}),
one finds the polarized coherent state from Eq.~(\ref{eq:cohere_state})
written as
\begin{equation}
\mathcal{J}\left|\alpha(\theta,\phi)\right\rangle =e^{-|\alpha|^{2}/2}\sum^{\infty}_{n=0}\sum^{n}_{k=0}\frac{\tilde{\alpha}^{n}e^{2ik\varphi}}{2^{n}\sqrt{(n-k)!k!}}\left|\circlearrowleft^{n-k},\circlearrowright^{k}\right\rangle \label{eq:Jones_coh}
\end{equation}
where $\tilde{\alpha}=\alpha(\cos\theta+e^{i(\phi-2\varphi)}\sin\theta)$
denotes the polarized displacement. This state corresponds to the
lower reference beam in Fig.~\ref{fig:interferometer_model}.

The effect of polarization on the left DPHS then follows as

\begin{multline}
\mathcal{J}D(\bm{\alpha}_{\theta,\phi})\left|\circlearrowleft\right\rangle =e^{-|\alpha|^{2}/2}\sum^{\infty}_{n=0}\sum^{n}_{k=0}\frac{\tilde{\alpha}^{n}e^{2ik\varphi}}{2^{n}\sqrt{(n-k)!k!}}\\
\times\left[\sqrt{n-k+1}\left|\circlearrowleft^{n-k+1},\circlearrowright^{k}\right\rangle -\alpha^{\ast}\cos\theta\left|\circlearrowleft^{n-k},\circlearrowright^{k}\right\rangle \right].\label{eq:polar_DPHS}
\end{multline}
The right DPHS after polarization has a similar formula except $\sqrt{n-k+1}\left|\circlearrowleft^{n-k+1},\circlearrowright^{k}\right\rangle $
being replaced by $\sqrt{k+1}\left|\circlearrowleft^{n-k},\circlearrowright^{k+1}\right\rangle $.
These formulas corresponds to the upper signal beam in Fig.~\ref{fig:interferometer_model}.

\subsection{Stokes parameter}

When the signal beam and the reference beam combine at the 50:50 beam
splitter, they interfere in the expanded Hilbert space $\mathcal{H}=\mathcal{H}_{\mathrm{sig}}\times\mathcal{H}_{\mathrm{ref}}$
of signal and reference. A photo detector at one end of the outputs
measures power level proportional to the photon count. In other words,
it measures essentially the zeroth Stokes parameter $S_{0}$ as the
state average $\left\langle \Sigma_{0}\right\rangle $ of the operator
$\Sigma_{0}=c^{\dag}c+\bar{c}^{\dag}\bar{c}$ on $\mathcal{H}$, where
$\{c,c^{\dag}\}$ (resp. $\{\bar{c},\bar{c}^{\dag}\})$ denotes the
annihilation and creation operator pair of the left-helical (resp.
right-helical) photons in the measured output beam.

Letting the signal beam photons associate with the operators $a$
and $\bar{a}$ (both helical directions) as in Eqs.~(\ref{eq:left_DPHS})-(\ref{eq:right_DPHS})
and the reference beam photons with the operators $b$ and $\bar{b}$,
the beam splitting process establishes the relations $c=(a+b)/\sqrt{2}$
and $\bar{c}=(\bar{a}+\bar{b})/\sqrt{2}$ between the output and the
input. Consequently, the photon-counting operator $\Sigma_{0}$ can
be written as
\begin{align}
\Sigma_{0} & =\frac{1}{2}(a^{\dag}a+\bar{a}^{\dag}\bar{a}+b^{\dag}b+\bar{b}^{\dag}\bar{b})\label{eq:Stokes_def}\\
 & \qquad\qquad+\frac{1}{2}(a^{\dag}b+ab^{\dag}+\bar{a}^{\dag}\bar{b}+\bar{a}\bar{b}^{\dag})\nonumber 
\end{align}
where the first term accounts for the incident photon number from
both inputs and the second for the photonic interference.

First, we consider the AF interference of two exactly identical coherent
states, which would serve as a baseline standard for those interferences
between DPHS and coherent states. In other words, using Eq.~(\ref{eq:Jones_coh})
for both the upper signal beam and the lower reference beam, we find
(Cf. App.~\ref{sec:appendix c}) that
\begin{align}
S_{0,\mathrm{coh}} & =|\alpha|^{2}\left(1+\sin2\theta\cos(\phi-2\varphi)\right)\label{eq:Stokes_coh}\\
 & \qquad\exp\left\{ -|\alpha|^{2}(1-\sin2\theta\cos(\phi-2\varphi))\right\} \nonumber 
\end{align}
The incident photon number term and the interference term of Eq.~(\ref{eq:Stokes_def})
has contributed equally in Eq.~(\ref{eq:Stokes_coh}), as expected
from the exact symmetry between the upper and the lower beam compositions.
The expression implies the dependence of interference fringes on the
polarizing angles $(\theta,\phi)$ of the incident (elliptically)
polarized beam as well as the orientation angle $\varphi$ of the
polarizers, coinciding with classical Arago-Fresnel interferometry.
In particular, parameter $S_{0,\mathrm{coh}}$ obtains the maximal
value $2|\alpha|^{2}$ when the tilting angle of the ellipse major
axis aligns with the polarizer orientation at $\phi=2\varphi$ and
the polarization angle $\theta=\pi/4$. In other words, when properly
oriented, the polarizers let through two linearly polarized beams
and let them constructively interfere such that the output power is
just the sum of the input powers.

The more interesting cases are those where the state of the upper
beam is a DPHS parametrically down-converted from a coherent state
with a single unknown photon, as shown in Fig.~\ref{fig:interferometer_model}.
Using Eq.~(\ref{eq:polar_DPHS}) for left DPHS and its complement
for right DPHS (Cf. App.~\ref{sec:appendix d}), we find that 
\begin{align}
 & S_{0,L\mathrm{DPHS}}=\Bigl\{\frac{1}{2}+|\alpha|^{2}\Bigl[\frac{1}{2}+\sin2\theta\cos(\phi-2\varphi)+\sin^{2}\theta\Bigr]\nonumber \\
 & +\frac{|\alpha|^{4}}{4}[1-\sin^{2}2\theta\cos^{2}(\phi-2\varphi)]\Bigr\} e^{-|\alpha|^{2}(1-\sin2\theta\cos(\phi-2\varphi))}\label{eq:Stokes_LDPHS}\\
 & S_{0,R\mathrm{DPHS}}=\Bigl\{\frac{1}{2}+|\alpha|^{2}\Bigl[\frac{1}{2}+\sin2\theta\cos(\phi-2\varphi)+\cos^{2}\theta\Bigr]\nonumber \\
 & +\frac{|\alpha|^{4}}{4}[1-\sin^{2}2\theta\cos^{2}(\phi-2\varphi)]\Bigr\} e^{-|\alpha|^{2}(1-\sin2\theta\cos(\phi-2\varphi))}\label{eq:Stokes_RDPHS}
\end{align}
While the exponential part is retained from Eq.~(\ref{eq:Stokes_coh}),
the leading factor now bears the extra constant and $|\alpha|^{4}$
coefficients characteristic of the DPHS. Furthermore, the $|\alpha|^{2}$
coefficient now has the extra $\sin^{2}\theta$ and $\cos^{2}\theta$
proportion terms that diffenentiate the left and right DPHS. In other
words, a particular elliptically polarized carrier beam breaks the
left-right helicity symmetry. The fringe produced by the AF interference
signifies the particular helicity. 

\section{Interference fringe and correlation of photon helicities}

\begin{figure*}
\includegraphics[bb=20bp 0bp 1020bp 303bp,clip,width=15cm]{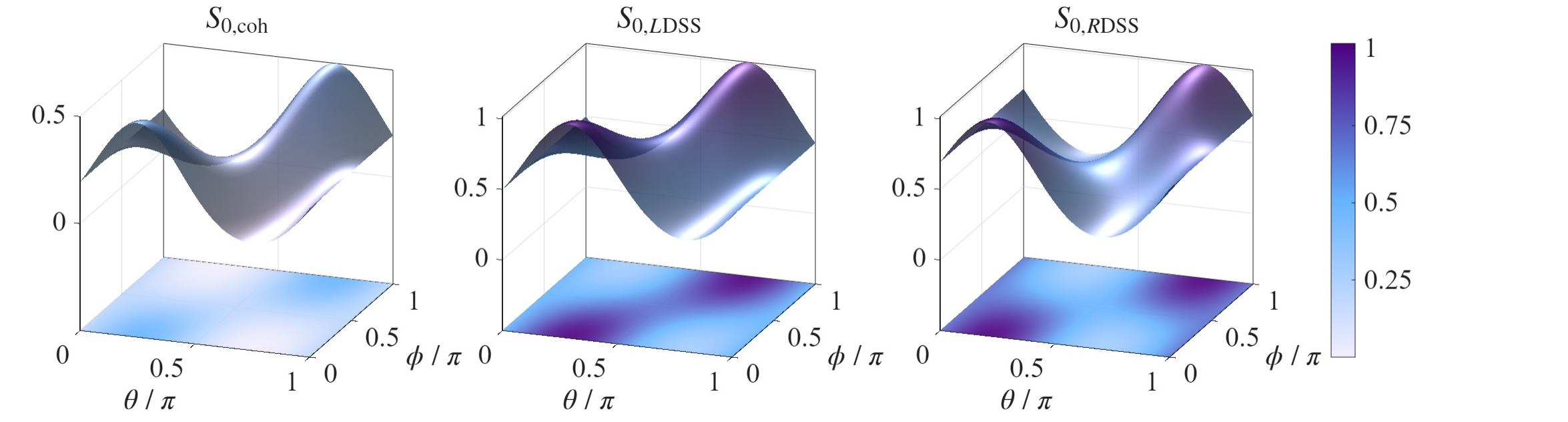}

\caption{(a) Surface plot overlaid with underneath contour of the Stokes parameter
$S_{0}$ for Arago-Fresnel (AF) self-interference of a coherent-state
beam (acting as both the signal and the reference beams) of polarizing
angles $\theta$ and $\phi$ . The plot aligns with our expectations
with polarized beam behaviors in classical AF laws and serves as a
reference baseline. (b) The same plot where the signal beam is in
a left displaced photon helicity states (DPHS). (c) The same plot
where the signal beam is in a right DPHS. In all plots, the polarizer
orientation is set to $\varphi=0$ and the displacement of the coherent
state is set to $\alpha=1/2$.~}\label{fig:Stokes_weak_alpha}
\end{figure*}

The Stokes parameters $S_{0}$ computed in last section suggest that
the output photon count from the interferometer varies in specific
patterns according to the angles of the elliptic polarization (described
by Eq.~(\ref{eq:arb_spin})) and the alignment angle of the polarizer
(described by Eq.~(\ref{eq:Jones})). As a baseline standard, Fig.~\ref{fig:Stokes_weak_alpha}(a)
plots the $S_{0}$ according to Eq.~(\ref{eq:Stokes_coh}) for a
polarized coherent state interfering with itself, where the polarizer
orientation is set to $\varphi=0$ and the coherent displacement of
the reference beam is set to $1/2$. Coinciding with the classical
AF laws, the magnitude of $S_{0}$ fringes maximizes to $1/2$ when
the beam is linearly polarized along the diagonal axis at $\theta=\pi/4$
(resp. $3\pi/4$) and $\phi=0$ (resp. $\pi$), demonstrating constructive
inferences. It minimizes to $0$ when the beam is linearly diagonalized
along the anti-diagonal axis at $\theta=3\pi/4$ (resp. $\pi/4$)
and $\phi=0$ (resp. $\pi$) for destructive inference.

In contrast, as illustrated in Figs.~\ref{fig:Stokes_weak_alpha}(b)
and (c) where $\alpha$ is set to $1/2$, both the left and the right
DPHS display elevated non-zero $S_{0}$ across the full ranges of
$\theta$ and $\phi$, due to the offset by the constant coefficient
$1/2$. Between the two DPHS, however, their $S_{0}$ variations about
$\theta$ and $\phi$ are distinct, as emphasized by the overlay contours
in the figure. For instance, though both DPHS reach a maxmimum $S_{0}$
magnitude of l at $\phi=0$, the right DPHS obtains it at $\theta=\frac{1}{2}\tan^{-1}(\frac{1}{2})$
while the left DPHS at $\theta=\frac{\pi}{2}-\frac{1}{2}\tan^{-1}(\frac{1}{2})$.
In addition, at $\theta=0$, $S_{0,R\mathrm{DPHS}}$ differs from
$S_{0,L\mathrm{DPHS}}$ by $|\alpha|^{2}e^{-|\alpha|^{2}}$ . These
distinguishing features imply that using a left circular polarized
($\theta=0$) reference distinguishes the spin direction of an unknown
incident photon. 

\begin{figure*}
\includegraphics[bb=25bp 0bp 1005bp 303bp,clip,width=15cm]{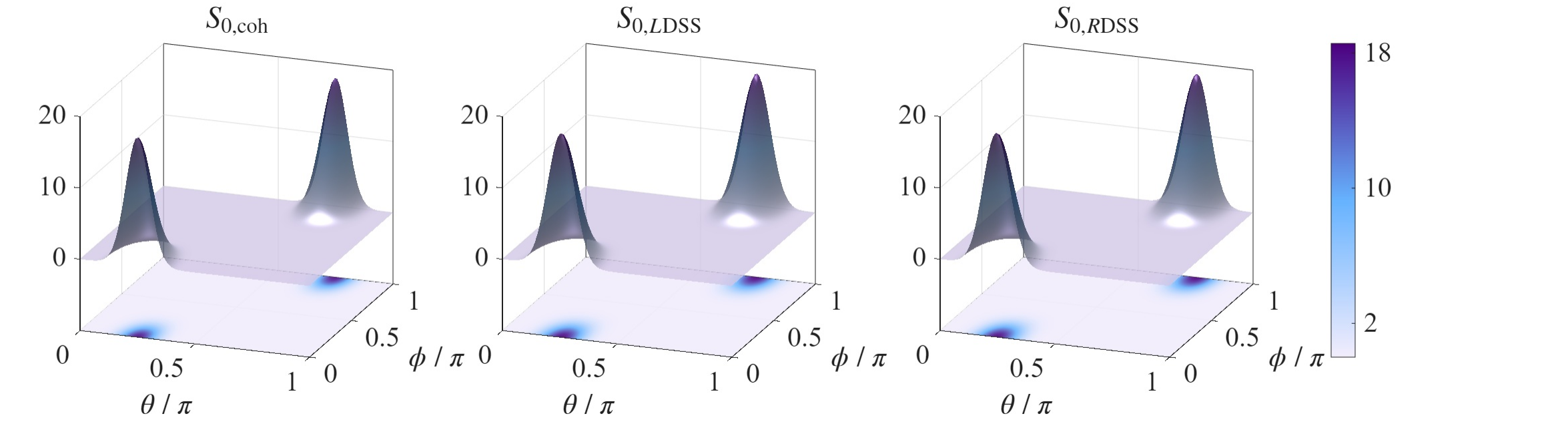}

\caption{Surface plots overlaid with underneath contours of the Stokes parameters
(a) $S_{0,\mathrm{coh}}$, (b) $S_{0,L\mathrm{DPHS}}$, and (c) $S_{0,R\mathrm{DPHS}}$,
i.e. the same plots as in Fig.~\ref{fig:Stokes_weak_alpha}, but
with displacement $\alpha=3$ to indicate a stronger powered reference
beam.~}\label{fig:Stokes_strong_alpha}

\end{figure*}

Finally, we remark that the non-classicality of this mesoscopic light
state plays a key role in producing the distinct interference patterns.
On one hand, at the microscopic limit of single photon states with
vanishing displacement (i.e. $\alpha\to0$), the Stokes parameters
for the two DPHS degenerate to a non-distinguishing constant $S_{0,L\mathrm{DPHS}}=S_{0,R\mathrm{DPHS}}\to1/2$.
On the other, at the macroscopic limit of strong light intensity where
the displacement $\alpha\to\infty$, the two parameters also degenerate
when the $|\alpha|^{4}$ term in dominates over the $|\alpha|^{2}$
term after $|\alpha|>1$. As shown in Fig.~\ref{fig:Stokes_strong_alpha},
for $\alpha=3$, the variations of $S_{0}$ of the DPHS are already
almost indistinguishable from that of the coherent state.

\section{Conclusions}

We introduced displaced photon helicity states, which are examined
through a binomial analysis on the gauge-invariant helicity basis
to infer their interferences with a reference coherent state. We establish
this quantum interometry by effecting a displacement on the helicity
of photon through a special parametric down conversion, followed by
a traditional Arago-Fresnel type interferometer. Through the computation
of Stokes parameters, the fringe is found to depend on the photon
helicity as well as the norm and phase of the displacement. Thereby,
the classical interference phenomenon is extended into the quantum
regime, whence the quantum AF interferometer serves as a single-photon
helicity detector.

\section*{Acknowledgments}

H. I. acknowledges the support by FDCT of Macau under grants 0179/2023/RIA3
and 0134/2024/AFJ, and by the Guangdong Provincial Quantum Science
Strategic Initiative (Grants No. GDZX2203001, GDZX2403001).

\appendix

\section{Coherent state from photon helicities}\label{sec:appendix a}

To show the identities of Eqs.~(\ref{eq:left_DPHS})-(\ref{eq:right_DPHS})
of the main text, we first rearrange terms in the displacement operator
as
\begin{align}
D(\bm{\alpha}_{\theta,\phi}) & =\exp\left\{ \alpha(a^{\dag}\cos\theta+\bar{a}^{\dag}e^{i\phi}\sin\theta)\right.\nonumber \\
 & \qquad\left.-\alpha^{\ast}(a\cos\theta+\bar{a}e^{-i\phi}\sin\theta)\right\} \nonumber \\
 & =\exp\left\{ (\alpha a^{\dag}-\alpha^{\ast}a)\cos\theta\right\} \nonumber \\
 & \qquad\exp\left\{ (\alpha e^{i\phi}\bar{a}^{\dag}-\alpha^{\ast}e^{-i\phi}\bar{a})\sin\theta\right\} 
\end{align}
of the two independent, thus commutative, helicity modes associated
with the operators $a$ and $\bar{a}$. Hence, exerting the displacement
$\alpha(\theta,\phi)$ on the vacuum $\left|0\right\rangle $ is equivalent
to exerting a complementary pair of displacements: $\alpha\cos\theta$
on the left-helical mode and $\alpha e^{i\phi}\sin\theta$ on the
right-helical mode on the double vaccum $\left|0,0\right\rangle $.
Given the commutativity $[a,\bar{a}^{\dag}]=0$, the usual unitarity,
i.e. 
\begin{equation}
D^{\dag}(\bm{\alpha}_{\theta,\phi})D(\bm{\alpha}_{\theta,\phi})=I\quad D^{\dag}(\bm{\alpha}_{\theta,\phi})=D(-\bm{\alpha}_{\theta,\phi}),
\end{equation}
is observed.

Then, using this unitarity, we find for the left DPHS
\begin{align}
D & (\bm{\alpha}_{\theta,\phi})\left|\circlearrowleft\right\rangle =D(\bm{\alpha}_{\theta,\phi})a^{\dag}\left|0,0\right\rangle \nonumber \\
 & =\left[D^{\dag}(-\bm{\alpha}_{\theta,\phi})a^{\dag}D(-\bm{\alpha}_{\theta,\phi})\right]D(\bm{\alpha}_{\theta,\phi})\left|0,0\right\rangle .
\end{align}
For displacing $a^{\dagger}$ in the square bracket, we invoke Baker-Campbell-Hausdorff
(BCH) expansion, for which the exponentials about $\{\bar{a},\bar{a}^{\dag}\}$
are cancelled out, to be left with
\begin{align}
 & D^{\dag}(-\bm{\alpha}_{\theta,\phi})a^{\dag}D(-\bm{\alpha}_{\theta,\phi})\nonumber \\
= & \exp\{(\alpha a^{\dag}-\alpha^{\ast}a)\cos\theta\}a^{\dagger}\exp\{(\alpha a^{\dag}-\alpha^{\ast}a)\cos\theta\}\nonumber \\
= & a^{\dagger}-\alpha^{\ast}\cos\theta\label{eq:s_disp_create}
\end{align}

For the factor $D_{\theta,\phi}(\alpha)\left|0\right\rangle $, we
use the BCH formula again to obtain the coherent state (Eq.~(\ref{eq:cohere_state})
of the main text) in terms of the binomial distribution of photon
helicities:
\begin{align}
 & \left|\alpha(\theta,\phi)\right\rangle =D(\bm{\alpha}_{\theta,\phi})\left|0,0\right\rangle \nonumber \\
= & \exp\left\{ (\alpha a^{\dag}-\alpha^{\ast}a)\cos\theta+(\alpha e^{i\phi}\bar{a}^{\dag}-\alpha^{\ast}e^{-i\phi}\bar{a})\sin\theta\right\} \left|0,0\right\rangle \nonumber \\
= & \exp\left\{ -\frac{|\alpha|^{2}}{2}\right\} \exp\left\{ \alpha\cos\theta a^{\dag}\right\} \exp\left\{ \alpha e^{i\phi}\sin\theta\bar{a}^{\dag}\right\} \left|0,0\right\rangle \nonumber \\
= & e^{-|\alpha|^{2}/2}\sum^{\infty}_{n=0}\frac{1}{n!}\left[(\alpha\cos\theta)a^{\dag}+(\alpha e^{i\phi}\sin\theta)\bar{a}^{\dag}\right]^{n}\left|0,0\right\rangle \nonumber \\
= & e^{-|\alpha|^{2}/2}\sum^{\infty}_{n=0}\frac{\alpha^{n}}{n!}\sum^{n}_{k=0}\binom{n}{k}e^{ik\phi}\cos^{n-k}\theta\sin^{k}\theta\nonumber \\
 & \qquad\qquad a^{\dag(n-k)}\bar{a}^{\dag k}\left|0,0\right\rangle \nonumber \\
= & e^{-|\alpha|^{2}/2}\sum^{\infty}_{n=0}\sum^{n}_{k=0}\frac{\alpha^{n}e^{ik\phi}}{\sqrt{(n-k)!k!}}\cos^{n-k}\theta\sin^{k}\theta\left|\circlearrowleft^{n-k},\circlearrowright^{k}\right\rangle \label{eq:s_cohere_state}
\end{align}

Combining Eqs.~(\ref{eq:s_disp_create})-(\ref{eq:s_cohere_state})
above, the identity (Eq.~(\ref{eq:left_DPHS})) of the main text
is recovered. Substituting $a$ by $\bar{a}$ and $\cos\theta$ by
$e^{-i\phi}\sin\theta$ in the above argument leads to the identity
(Eq.~(\ref{eq:right_DPHS})) for the right DPHS.

\section{Polarized states through Jones matrix}\label{sec:appendix b}

Applying the single-photon Jones matrix $\hat{\mathcal{J}}$ on the
single-photon state (Eq.~(\ref{eq:arb_spin}) of the main text) leads
to
\begin{align}
\hat{\mathcal{J}}\left|\mathbf{1}_{\theta,\phi}\right\rangle  & =\frac{1}{2}\cos\theta(\left|\circlearrowleft\right\rangle +e^{2i\varphi}\left|\circlearrowright\right\rangle )\nonumber \\
 & +\frac{1}{2}e^{i\phi}\sin\theta(e^{-2i\varphi}\left|\circlearrowleft\right\rangle +\left|\circlearrowright\right\rangle )\nonumber \\
 & =\frac{1}{2}\left[\cos\theta+e^{i(\phi-2\varphi)}\sin\theta\right]\left(\left|\circlearrowleft\right\rangle +e^{2i\varphi}\left|\circlearrowright\right\rangle \right).
\end{align}
Then taking the tensor product for $n$ such polarized photons, one
obtains the effect of Jones matrix on a Fock state as
\begin{align}
\hat{\mathcal{J}}^{\otimes n}\left|\mathbf{n}_{\theta,\phi}\right\rangle  & =\frac{1}{2^{n}}\left[\cos\theta+e^{i(\phi-2\varphi)}\sin\theta\right]^{n}\nonumber \\
 & \sum^{n}_{k=0}\binom{n}{k}e^{2ik\varphi}\left|\circlearrowleft^{n-k},\circlearrowright^{k}\right\rangle .
\end{align}
Consequently, the coherent state after applying the Jones matrix becomes
\begin{align}
 & \mathcal{J}\left|\alpha(\theta,\phi)\right\rangle =e^{-|\alpha|^{2}/2}\sum^{\infty}_{n=0}\frac{\alpha^{n}}{n!}\mathcal{J}^{\otimes n}\left|n(\theta,\phi)\right\rangle )\nonumber \\
 & =e^{-|\alpha|^{2}/2}\sum^{\infty}_{n=0}\left(\frac{\alpha}{2}\right)^{n}\left[\cos\theta+e^{i(\phi-2\varphi)}\sin\theta\right]^{n}\nonumber \\
 & \qquad\sum^{n}_{k=0}\frac{e^{2ik\varphi}}{\sqrt{(n-k)!k!}}\left|\circlearrowleft^{n-k},\circlearrowright^{k}\right\rangle \nonumber \\
 & =e^{-|\alpha|^{2}/2}\sum^{\infty}_{n=0}\left(\frac{\tilde{\alpha}}{2}\right)^{n}\sum^{n}_{k=0}\frac{e^{2ik\varphi}}{\sqrt{(n-k)!k!}}\left|\circlearrowleft^{n-k},\circlearrowright^{k}\right\rangle \label{eq:s_Jones_coher}
\end{align}
where $\tilde{\alpha}=\alpha(\cos\theta+e^{i(\phi-2\varphi)}\sin\theta)$
is the polarized displacement. This is Eq.~(\ref{eq:Jones_coh})
of the main text and Eq.~(\ref{eq:polar_DPHS}) immediately follows
by applying $(a^{\dagger}-\alpha^{\ast}\cos\theta)$ to $n$-photon
state $\left|\circlearrowleft^{n-k},\circlearrowright^{k}\right\rangle $.

\section{Computing Stokes parameter for baseline coherent state}\label{sec:appendix c}

To streamline the computation of the Stokes parameter $S_{0}=\left\langle \Sigma_{0}\right\rangle $
where $\Sigma_{0}$ is defined by Eq.~(\ref{eq:Stokes_def}) of the
main text, we separate this Stokes operator into two parts 
\begin{align}
\Sigma^{\mathcal{N}}_{0}= & \frac{1}{2}\left(a^{\dag}a+\bar{a}^{\dag}\bar{a}+b^{\dag}b+\bar{b}^{\dag}\bar{b}\right),\\
\Sigma^{\mathcal{I}}_{0}= & \frac{1}{2}\left(a^{\dag}b+ab^{\dag}+\bar{a}^{\dag}\bar{b}+\bar{a}\bar{b}^{\dag}\right),
\end{align}
to denote respectively the total incident photon numbers and the interference
across two beams. We first consider the case where the signal beam
(associated with $a$ and $\bar{a}$ for the two helical orientations)
is identical to the reference beam (associated with $b$ and $\bar{b}$
for the two helical orientations), both decribed by the polarized
coherent state Eq.~(\ref{eq:s_Jones_coher}). Then, in the expanded
Hilbert space $\mathcal{H}=\mathcal{H}_{\mathrm{sig}}\times\mathcal{H}_{\mathrm{ref}}$,
the joint product state $\mathcal{J}\left|\alpha(\theta,\phi)\right\rangle \otimes\mathcal{J}\left|\alpha(\theta,\phi)\right\rangle $
is expanded under the Fock eigenvectors $\left|\circlearrowleft^{n-k},\circlearrowright^{k};\circlearrowleft^{n'-k'},\circlearrowright^{k'}\right\rangle $,
where the un-primed helicities refer to the upper signal and the primed
to the lower reference. With such a basis, $\left\langle \Sigma^{\mathcal{N}}_{0}\right\rangle $
concerns with the matrix elements
\begin{multline}
\left\langle \circlearrowleft^{m-l},\circlearrowright^{l};\circlearrowleft^{m'-l'},\circlearrowright^{l'}\right|\left[a^{\dag}a+\bar{a}^{\dag}\bar{a}+b^{\dag}b+\bar{b}^{\dag}\bar{b}\right]\\
\left|\circlearrowleft^{n-k},\circlearrowright^{k};\circlearrowleft^{n'-k'},\circlearrowright^{k'}\right\rangle \\
=(n+n')\delta_{m-l,n-k}\delta_{l,k}\delta_{m'-l',n'-k'}\delta_{l',k'}.\label{eq:s_num_Fock_matrix}
\end{multline}
With four-fold superposition expansions associated with the four index
pairs $\{m,l\}$, $\{m',l'\}$, $\{n,k\}$, and $\{n',k'\}$, we find
\begin{align}
 & \left\langle \Sigma^{\mathcal{N}}_{0}\right\rangle _{\mathrm{coh}}=\frac{e^{-2|\alpha|^{2}}}{2}\biggl\{\sum^{\infty}_{m,n=0}\sum^{\infty}_{m',n'=0}\left(\frac{\tilde{\alpha}}{2}^{\ast}\right)^{m+m'}\left(\frac{\tilde{\alpha}}{2}\right)^{n+n'}\nonumber \\
 & \qquad\sum^{m}_{l=0}\sum^{n}_{k=0}\frac{e^{2i(k-l)\varphi}}{\sqrt{(m-l)!l!(n-k)!k!}}\nonumber \\
 & \qquad\sum^{m'}_{l'=0}\sum^{n'}_{k'=0}\frac{e^{2i(k'-l')\varphi}}{\sqrt{(m'-l')!l'!(n'-k')!k'!}}(n+n')\nonumber \\
 & \qquad\qquad\delta_{m-l,n-k}\delta_{l,k}\delta_{m'-l',n'-k'}\delta_{l',k'}\biggr\}.
\end{align}

Collapsing the deltas leads to the formula
\begin{align}
\left\langle \Sigma^{\mathcal{N}}_{0}\right\rangle _{\mathrm{coh}}= & \frac{e^{-2|\alpha|^{2}}}{2}\sum^{\infty}_{n,n'=0}\left(\frac{|\tilde{\alpha}|^{2}}{4}\right)^{n+n'}\nonumber \\
 & \sum^{n}_{k=0}\frac{1}{(n-k)!k!}\sum^{n'}_{k'=0}\frac{1}{(n'-k')!k'!}(n+n')
\end{align}
where the summation of the factorials over $k$ and $k'$ can be binomially
contracted into $2^{n}$ and $2^{n'}$ such that
\begin{align}
\left\langle \Sigma^{\mathcal{N}}_{0}\right\rangle _{\mathrm{coh}} & =\frac{e^{-2|\alpha|^{2}}}{2}\sum^{\infty}_{n,n'=0}\frac{1}{n!n'!}\left(\frac{|\tilde{\alpha}|^{2}}{2}\right)^{n+n'}(n+n')\nonumber \\
 & =\frac{e^{-2|\alpha|^{2}}}{2}e^{|\tilde{\alpha}|/2}\biggl\{\sum^{\infty}_{n=0}\frac{1}{(n-1)!}\left(\frac{|\tilde{\alpha}|^{2}}{2}\right)^{n}\nonumber \\
 & \quad+\sum^{\infty}_{n'=0}\frac{1}{(n'-1)!}\left(\frac{|\tilde{\alpha}|^{2}}{2}\right)^{n'}\biggr\}.
\end{align}
Note that $n/n!$ at $n=0$ and $n'/n'!$ at $n'=0$ contribute zero-value
terms. We can thus shift the summation indices $n$ and $n'$ by $1$
($n\to n+1$) to find
\begin{equation}
\left\langle \Sigma^{\mathcal{N}}_{0}\right\rangle _{\mathrm{coh}}=\frac{|\tilde{\alpha}|^{2}}{2}e^{-2|\alpha|^{2}+|\tilde{\alpha}|}.
\end{equation}
Since $|\tilde{\alpha}|\leq|\alpha|$, the exponential factor above
is convergent.

For the second term $\left\langle \Sigma^{\mathcal{I}}_{0}\right\rangle $
of the Stokes parameter, consider the matrix elements
\begin{align}
 & \left\langle \circlearrowleft^{m-l},\circlearrowright^{l};\circlearrowleft^{m'-l'},\circlearrowright^{l'}\right|\left[a^{\dag}b+ab^{\dag}+\bar{a}^{\dag}\bar{b}+\bar{a}\bar{b}^{\dag}\right]\nonumber \\
 & \quad\left|\circlearrowleft^{n-k},\circlearrowright^{k};\circlearrowleft^{n'-k'},\circlearrowright^{k'}\right\rangle \nonumber \\
 & =\sqrt{(n-k+1)(n'-k')}\delta_{m-l,n-k+1}\delta_{m'-l',n'-k'-1}\nonumber \\
 & \:+\sqrt{(n-k)(n'-k'+1)}\delta_{m-l,n-k-1}\delta_{l,k}\delta_{m'-l',n'-k'+1}\delta_{l',k'}\nonumber \\
 & \:+\sqrt{(k+1)k'}\delta_{m-l,n-k}\delta_{l,k+1}\delta_{m'-l',n'-k'}\delta_{l',k'-1}\nonumber \\
 & \:+\sqrt{k(k'+1)}\delta_{m-l,n-k}\delta_{l,k-1}\delta_{m'-l',n'-k'}\delta_{l',k'+1}.
\end{align}
Similar in vein to the derivation for $\left\langle \Sigma^{\mathcal{N}}_{0}\right\rangle $,
after collapsing the deltas in the four-fold superposition expansions,
we obtain
\begin{align}
\left\langle \Sigma^{\mathcal{I}}_{0}\right\rangle _{\mathrm{coh}} & =\frac{e^{-2|\alpha|^{2}}}{2}\sum^{\infty}_{n,n'=0}\left(\frac{|\tilde{\alpha}|^{2}}{4}\right)^{n+n'+1}\nonumber \\
 & \quad\biggl\{\sum^{n}_{k=0}\frac{1}{(n-k)!k!}\sum^{n'}_{k'=0}\frac{1}{(n'-k')!k'!}\nonumber \\
 & \quad+\sum^{n}_{k=0}\frac{1}{(n-k)!k!}\sum^{n'}_{k'=0}\frac{1}{(n'-k')!k'!}\nonumber \\
 & \quad+\sum^{n}_{k=0}\frac{1}{(n-k)!k!}\sum^{n'+1}_{k'=1}\frac{1}{(n'-k'+1)!(k'-1)!}\nonumber \\
 & \quad+\sum^{n+1}_{k=1}\frac{1}{(n-k+1)!(k-1)!}\sum^{n'}_{k'=0}\frac{1}{(n'-k')!k'!}\biggr\},
\end{align}
where the summation indices are necessarily adjusted to accommodate
the coefficients containing $n,k,n'$, and $k'$ in front of deltas
that would make the leading terms vanish in the summations. Again,
forming binomial coefficients in these sums, we observe each term
in the braces are actually the same such that
\begin{align}
\left\langle \Sigma^{\mathcal{I}}_{0}\right\rangle _{\mathrm{coh}} & =\frac{e^{-2|\alpha|^{2}}}{2}\sum^{\infty}_{n,n'=0}\frac{4}{n!n'!}\left(\frac{|\tilde{\alpha}|^{2}}{4}\right)^{n+n'+1}2^{n+n'}\nonumber \\
 & =\frac{|\tilde{\alpha}|^{2}}{2}e^{-2|\alpha|^{2}+|\tilde{\alpha}|}.
\end{align}
Therefore, we find that $\left\langle \Sigma^{\mathcal{N}}_{0}\right\rangle $
and $\left\langle \Sigma^{\mathcal{I}}_{0}\right\rangle $ contribute
equally and the full Stokes parameter for AF interfering two identical
coherent state reads
\begin{align}
S_{0} & =|\tilde{\alpha}|^{2}e^{-2|\alpha|^{2}+|\tilde{\alpha}|}\nonumber \\
 & =\left(1+\sin2\theta\cos(\phi-2\varphi)\right)|\alpha|^{2}\nonumber \\
 & \quad\times e^{-|\alpha|^{2}\left(1-\sin2\theta\cos(\phi-2\varphi)\right)}.
\end{align}

\section{Computing Stokes parameter for DPHS}\label{sec:appendix d}

We now consider the main case of interfering a left DPHS with a coherent
state, letting the upper beam (associated with $a$ and $\bar{a}$)
be the DPHS and the lower beam (associated with $b$ and $\bar{b}$)
be the coherent state. Again, the derivations are similar to those
in Sec. 3 since the expansion coefficients are almost identical between
a polarized coherent state and a polarized DPHS after applying the
Jones matrix. The only difference is that, from Eq.~(\ref{eq:polar_DPHS})
of the main text, the polarized DPHS comprises two terms instead of
one: i) the displacement coefficient term, which does not add or subtract
photon numbers; and ii) the $a^{\dagger}$ term, which adds a photon
along the corresponding helicitiy. Therefore, we need to consider
three extra cases besides Eq.~(\ref{eq:s_num_Fock_matrix}) when
computing $\left\langle \Sigma^{\mathcal{N}}_{0}\right\rangle $:
1) the vector as the ket having one extra photon, 2) the dual vector
as the bra having one extra photon, and 3) both the bra and the ket
having one extra photon, i.e. 
\begin{align}
 & \left\langle \circlearrowleft^{m-l},\circlearrowright^{l};\circlearrowleft^{m'-l'},\circlearrowright^{l'}\right|\left[a^{\dag}a+\bar{a}^{\dag}\bar{a}+b^{\dag}b+\bar{b}^{\dag}\bar{b}\right]\nonumber \\
 & \quad\left|\circlearrowleft^{n-k+1},\circlearrowright^{k};\circlearrowleft^{n'-k'},\circlearrowright^{k'}\right\rangle \nonumber \\
 & =(n+n'+1)\delta_{m-l,n-k+1}\delta_{l,k}\delta_{m'-l',n'-k'}\delta_{l',k'},
\end{align}
\begin{align}
 & \left\langle \circlearrowleft^{m-l+1},\circlearrowright^{l};\circlearrowleft^{m'-l'},\circlearrowright^{l'}\right|\left[a^{\dag}a+\bar{a}^{\dag}\bar{a}+b^{\dag}b+\bar{b}^{\dag}\bar{b}\right]\nonumber \\
 & \quad\left|\circlearrowleft^{n-k},\circlearrowright^{k};\circlearrowleft^{n'-k'},\circlearrowright^{k'}\right\rangle \nonumber \\
 & =(n+n')\delta_{m-l+1,n-k}\delta_{l,k}\delta_{m'-l',n'-k'}\delta_{l',k'},
\end{align}
\begin{align}
 & \left\langle \circlearrowleft^{m-l+1},\circlearrowright^{l};\circlearrowleft^{m'-l'},\circlearrowright^{l'}\right|\left[a^{\dag}a+\bar{a}^{\dag}\bar{a}+b^{\dag}b+\bar{b}^{\dag}\bar{b}\right]\nonumber \\
 & \quad\left|\circlearrowleft^{n-k+1},\circlearrowright^{k};\circlearrowleft^{n'-k'},\circlearrowright^{k'}\right\rangle \nonumber \\
 & =(n+n'+1)\delta_{m-l,n-k}\delta_{l,k}\delta_{m'-l',n'-k'}\delta_{l',k'}.
\end{align}
With these formulas substituting into the expansions of the Stokes
operator, we find after collapsing the Kronecker deltas that
\begin{align}
 & \left\langle \Sigma^{\mathcal{N}}_{0}\right\rangle _{L\mathrm{DPHS}}=\frac{e^{-2|\alpha|^{2}}}{2}\sum^{\infty}_{n,n'=0}\left(\frac{|\tilde{\alpha}|^{2}}{4}\right)^{n+n'}\biggl\{|\alpha|^{2}\cos^{2}\theta\nonumber \\
 & \quad\sum^{n}_{k=0}\sum^{n'}_{k'=0}\frac{n+n'}{(n-k)!k!(n'-k')!k'!}\nonumber \\
 & -\frac{\cos\theta}{2}(\tilde{\alpha}^{\ast}\alpha+\tilde{\alpha}\alpha^{\ast})\sum^{n}_{k=0}\sum^{n'}_{k'=0}\frac{n+n'+1}{(n-k)!k!(n'-k')!k'!}\nonumber \\
 & +\sum^{n}_{k=0}\frac{n-k+1}{(n-k)!k!}\sum^{n'}_{k'=0}\frac{1}{(n'-k')!k'!}(n+n'+1)\biggr\}.
\end{align}
The expressions are similar to those in $\left\langle \Sigma^{\mathcal{I}}_{0}\right\rangle _{\mathrm{coh}}$.
So binomially contracting the finite sums leads to\begin{widetext}
\begin{align}
 & \left\langle \Sigma^{\mathcal{N}}_{0}\right\rangle _{L\mathrm{DPHS}}=\frac{e^{-2|\alpha|^{2}}}{4}\sum^{\infty}_{n,n'=0}\frac{1}{n!n'!}\left(\frac{|\tilde{\alpha}|^{2}}{2}\right)^{n+n'}\biggl\{2-\cos\theta(\tilde{\alpha}^{\ast}\alpha+\tilde{\alpha}\alpha^{\ast})\nonumber \\
 & \quad+\left[2|\alpha|^{2}\cos^{2}\theta-\cos\theta(\tilde{\alpha}^{\ast}\alpha+\tilde{\alpha}\alpha^{\ast})+2\right](n+n')+n(n-1)+nn'+2n\biggr\}.
\end{align}
Finally, forming exponentials from the infinite sums with necessary
shifting in summation indices, we have
\begin{align}
 & \left\langle \Sigma^{\mathcal{N}}_{0}\right\rangle _{L\mathrm{DPHS}}=\frac{1}{4}e^{-2|\alpha|^{2}+|\tilde{\alpha}|^{2}}\left\{ 2-\cos\theta(\tilde{\alpha}^{\ast}\alpha+\tilde{\alpha}\alpha^{\ast})+|\tilde{\alpha}|^{2}\left[2\cos^{2}\theta|\alpha|^{2}-\cos\theta(\tilde{\alpha}^{\ast}\alpha+\tilde{\alpha}\alpha^{\ast})+2\right]+\frac{|\tilde{\alpha}|^{4}}{4}+\frac{|\tilde{\alpha}|^{4}}{4}+|\tilde{\alpha}|^{2}\right\} \nonumber \\
 & \quad=\frac{1}{4}e^{-2|\alpha|^{2}+|\tilde{\alpha}|^{2}}\left[2|\alpha|^{2}|\tilde{\alpha}|^{2}\cos^{2}\theta-(1+|\tilde{\alpha}|^{2})(\tilde{\alpha}^{\ast}\alpha+\tilde{\alpha}\alpha^{\ast})\cos\theta+3|\tilde{\alpha}|^{2}+\frac{|\tilde{\alpha}|^{4}}{2}+2\right]\label{eq:s_DPHS_num}
\end{align}
where we have substituted the expression of the polarized displacement
$\tilde{\alpha}$ in the last equation.

Next, to compute the contribution from the interference term $\left\langle \Sigma^{\mathcal{I}}_{0}\right\rangle $,
we consider the matrix expansion of $a^{\dag}b+ab^{\dag}+\bar{a}^{\dag}\bar{b}+\bar{a}\bar{b}^{\dag}$
in the basis of $\mathcal{J}D_{\theta,\phi}(\alpha)\left|\circlearrowleft\right\rangle \otimes\mathcal{J}D_{\theta,\phi}(\alpha)\left|\circlearrowleft\right\rangle $
using Eq.~(\ref{eq:polar_DPHS}) of the mian text. Expanded in the
Fock bases for both the upper signal and the lower reference beams,
$\left\langle \Sigma^{\mathcal{I}}_{0}\right\rangle $ contains inner
products of the type
\begin{align}
 & \left\langle \circlearrowleft^{m-l+\mu},\circlearrowright^{l};\circlearrowleft^{m'-l'},\circlearrowright^{l'}\right|\left[a^{\dag}b+ab^{\dag}+\bar{a}^{\dag}\bar{b}+\bar{a}\bar{b}^{\dag}\right]\left|\circlearrowleft^{n-k+\nu},\circlearrowright^{k};\circlearrowleft^{n'-k'},\circlearrowright^{k'}\right\rangle \nonumber \\
 & =\sqrt{(n-k+\nu+1)(n'-k')}\delta_{m-l+\mu,n-k+\nu+1}\delta_{l,k}\delta_{m'-l',n'-k'-1}\delta_{l',k'}\nonumber \\
 & \quad+\sqrt{(n-k+\nu)(n'-k'+1)}\delta_{m-l+\nu,n-k+\mu-1}\delta_{l,k}\delta_{m'-l',n'-k'+1}\delta_{l',k'}\nonumber \\
 & \quad+\sqrt{(k+1)k'}\delta_{m-l+\mu,n-k+\nu}\delta_{l,k+1}\delta_{m'-l',n'-k'}\delta_{l',k'-1}+\sqrt{k(k'+1)}\delta_{m-l+\mu,n-k+\nu}\delta_{l,k-1}\delta_{m'-l',n'-k'}\delta_{l',k'+1}
\end{align}
where $\mu$ and $\nu$ take the value of either 0 or 1. When substituting
the above into the expansion and following the same routine used to
treat $\left\langle \Sigma^{\mathcal{N}}_{0}\right\rangle _{L\mathrm{DPHS}}$,
we obtain after grouping common factors of deltas
\begin{align}
\left\langle \Sigma^{\mathcal{I}}_{0}\right\rangle _{L\mathrm{DPHS}} & =\frac{e^{-2|\alpha|^{2}}}{2}\sum^{\infty}_{m,n,m',n'=0}\left(\frac{\tilde{\alpha}}{2}^{\ast}\right)^{m+m'}\left(\frac{\tilde{\alpha}}{2}\right)^{n+n'}\sum^{m,n,m',n'}_{l,k,l',k'=0}\frac{e^{2i(k+k'-l-l')\varphi}}{\sqrt{(m-l)!l!(n-k)!k!(m'-l')!l'!(n'-k')!k'!}}\nonumber \\
 & \biggl\{\biggl(\left[(|\alpha|^{2}\cos^{2}\theta+n-k+2)\delta_{m-l,n-k+1}-\alpha\cos\theta\sqrt{n-k+2}\delta_{m-l,n-k+2}\right]\sqrt{(n-k+1)(n'-k')}\delta_{m'-l'+1,n'-k'}\nonumber \\
 & \quad+\left[\left(|\alpha|^{2}\cos^{2}\theta+n-k+1\right)\delta_{m-l+1,n-k}-\alpha^{\ast}\cos\theta\sqrt{n-k-1}\delta_{m-l+2,n-k}\right]\sqrt{(n-k)(n'-k'+1)}\delta_{m'-l',n'-k'+1}\nonumber \\
 & \quad-\left[\alpha\sqrt{n'-k'+1}\delta_{m'-l',n'-k'+1}+\alpha^{\ast}\sqrt{n'-k'}\delta_{m'-l',n'-k'-1}\right](n-k+1)\cos\theta\delta_{m-l,n-k}\biggr)\delta_{l,k}\delta_{l',k'}\nonumber \\
 & +\left[(|\alpha|^{2}\cos^{2}\theta+n-k+1)\delta_{m-l,n-k}-\alpha\cos\theta\sqrt{n-k+1}\delta_{m-l,n-k+1}-\alpha^{\ast}\cos\theta\sqrt{n-k}\delta_{m-l+1,n-k}\right]\delta_{m'-l',n'-k'}\nonumber \\
 & \qquad\times\left[\sqrt{(k+1)k'}\delta_{l,k+1}\delta_{l',k'-1}+\sqrt{k(k'+1)}\delta_{l,k-1}\delta_{l',k'+1}\right]\biggr\}.
\end{align}
After collapsing the deltas and simplifying the expression, we find
\begin{equation}
\left\langle \Sigma^{\mathcal{I}}_{0}\right\rangle _{\mathrm{DPHS}}=\frac{1}{4}e^{-2|\alpha|^{2}+|\tilde{\alpha}|^{2}}\left[2|\alpha|^{2}|\tilde{\alpha}|^{2}\cos^{2}\theta-(1+|\tilde{\alpha}|^{2})(\alpha\tilde{\alpha}^{\ast}+\alpha^{\ast}\tilde{\alpha})\cos\theta+\frac{|\tilde{\alpha}|^{4}}{2}+3|\tilde{\alpha}|^{2}\right].
\end{equation}
Combining the contribution of the interference term with that of the
photon number term in Eq.~(\ref{eq:s_DPHS_num}), we obtain finally
\begin{align}
\left\langle \Sigma_{0}\right\rangle _{L\mathrm{DPHS}} & =\frac{1}{2}e^{-2|\alpha|^{2}+|\tilde{\alpha}|^{2}}\left[2|\alpha|^{2}|\tilde{\alpha}|^{2}\cos^{2}\theta-(1+|\tilde{\alpha}|^{2})\cos\theta(\tilde{\alpha}^{\ast}\alpha+\tilde{\alpha}\alpha^{\ast})+\frac{|\tilde{\alpha}|^{4}}{2}+3|\tilde{\alpha}|^{2}+1\right]\nonumber \\
=\frac{1}{4} & e^{-|\alpha|^{2}\left(1-\sin2\theta\cos(\phi-2\varphi)\right)}\biggl\{2+2|\alpha|^{2}\left[1+2\sin^{2}\theta+2\sin2\theta\cos(\phi-2\varphi)\right]+|\alpha|^{4}\left[1-\sin^{2}2\theta\cos^{2}(\phi-2\varphi)\right]\biggr\}.\label{eq:s_LDPHS}
\end{align}

The exact computation routine can be carried out for computing the
Stokes parameter for the right DPHS. Furthermore, observing the symmetry
in the expressions of Eqs.~(\ref{eq:left_DPHS}) and (\ref{eq:right_DPHS})
of the main text for the two DPHS, one can simply replace each $\cos\theta$
in Eq.~(\ref{eq:s_LDPHS}) and $\tilde{\alpha}$ with $\sin\theta$
to find equally 
\begin{align}
\left\langle \Sigma_{0}\right\rangle _{R\mathrm{DPHS}} & =\frac{1}{2}e^{-2|\alpha|^{2}+|\widetilde{\alpha}|^{2}}\left[2|\alpha|^{2}|\widetilde{\alpha}|^{2}\sin^{2}\theta-(1+|\widetilde{\alpha}|^{2})\sin\theta(\alpha^{*}\widetilde{\alpha}e^{-i(\phi-2\varphi)}+\mathrm{h.c.})+\frac{|\widetilde{\alpha}|^{4}}{2}+3|\widetilde{\alpha}|^{2}+1\right]\nonumber \\
=\frac{1}{4} & e^{-|\alpha|^{2}\left(1-\sin2\theta\cos(\phi-2\varphi)\right)}\biggl\{2+2|\alpha|^{2}\left[1+2\cos^{2}\theta+2\sin2\theta\cos(\phi-2\varphi)\right]+|\alpha|^{4}\left[1-\sin^{2}2\theta\cos^{2}(\phi-2\varphi)\right]\biggr\}.
\end{align}
\end{widetext}

\end{document}